\documentstyle[12pt,osa,aps,epsfig]{revtex}
\tighten

\newcommand{\rhog}{\mbox{$\phi$}}
\newcommand{\rhoe}{\mbox{$\rho$}}
\newcommand{\uf}{\mbox{$u$}}
\newcommand{\hc}{\mbox{$H$}}
\newcommand{\as}{\mbox{$\epsilon$}}
\newcommand{\ji}{\mbox{$j_{\infty}$}}
\newcommand{\gi}{\mbox{$g_{\infty}$}}
\newcommand{\ui}{\mbox{$u_{\infty}$}}

\begin{document}

\title{Diffusive Growth of a Polymer Layer by {\em In-Situ} Polymerization}

\author{J.~P. Wittmer $^{1,2}$,
M.~E. Cates $^1$,
A.~Johner$^3$,
M.~S. Turner$^2$}

\date{10 February 1996}
\setcounter{page}{0}
\maketitle

\vspace*{1cm}

$^1$ The University of Edinburgh, Dept. of Physics \& Astronomy,
King's Buildings, \\ Mayfield Road, Edinburgh EH9 3JZ, UK\\

$^2$ Cavendish Laboratory, Madingley Road, Cambridge CB3 0HE, UK\\

$^3$ Institut Charles Sadron, 6 rue Boussingault,
67083 Strasbourg, France

\vspace{1cm}
\begin{abstract}
We consider the growth of a polymer layer
on a flat surface in a good solvent by {\em in-situ} polymerization. This is
viewed
as a modified form of diffusion-limited aggregation without branching.
We predict theoretically the formation of
a pseudo-brush with density $\rhog(z) \propto z^{-2/3}$
and characteristic height $\hc \propto t^{3}$.
These results are found by combining a mean-field treatment of the
diffusive growth (marginally valid in three dimensions)
with a scaling theory (Flory exponent $\nu =3/5$) of the growing
polymers. We confirm their validity by Monte Carlo simulations.

\end{abstract}

\thispagestyle{empty}
\begin{center}
\end{center}

\vspace{1.5cm}


Polymer brushes (layers of linear polymer chains
end-grafted to a surface) find use in the
stabilization of colloidal particles in solution
~\cite{htl}.
One way to form these layers is
by adsorption of pre-existing polymers having a functionalised group
on one end, but this process turns out to be extremely slow for high
grafting densities~\cite{johner2}.
An alternative is the polymerization of a brush {\em in situ};
brushes formed this way have been reported, but not characterized in
detail
\cite{vidal}.
In this paper we study an idealized model of in-situ growth of a layer,
in which a
flat impenetrable surface is initially densely covered with
initiator sites. From these seeds, linear {unbranched} chains can
grow by irreversible polymerization, with
monomers added only at chain ends.
The chains so formed are treated as {\it flexible coils}: they can relax their
structure in response to excluded-volume interactions with their
neighbours. This process sets a dynamical time scale;
we restrict attention
to slow enough growth, such that
local equilibrium is maintained at all times.
We also assume a constant diffusivity for the free monomers
and a constant,
infinitesimal flux of incoming random walkers.
(For a finite, or a time-dependent, incident
flux, our
description breaks down, as discussed later.)

The above-formulated problem of {\em in situ} polymerization
is closely related to that of DLA
(Diffusion-Limited Aggregation \cite{wittensander}) without branching
\cite{rossi,meakin,cates,krug}.
There is no satisfactory mean-field fomulation for
ordinary DLA \cite{wittensander,ball}, but a continuum field
approach does exist to describe the growth of a forest of needles
by diffusive deposition, with adsorption on the needle-tips only
~\cite{cates}. It was found by simulation \cite{rossi} that the mean-field
calculation
is rather accurate in $d=3$ dimensions, and later realized
\cite{krug} that this is the upper critical dimension for the problem,
so that only logarithmic corrections to the mean-field
results should apply.
In what follows, we generalize the needle growth model to a much wider
class of processes, in which there is an arbitrary
power-law exponent $\nu$ defined by $\phi(z) \sim \xi(z)^{1/\nu-d}$
where $\phi(z)$ is the density at height $z$
and $\xi(z)$ the associated transverse separation between
linear growing objects. (This is the usual relation, for an object of
fractal dimension $1/\nu$, between the density and a ``blob size"
which is locally given by the separation $\xi(z)$.)
For in-situ polymerization in a good solvent, $\nu$
is simply the usual Flory exponent $\nu= 3/5$ \cite{degennesbook}.
In this case, we predict the formation of a novel
power-law ``pseudobrush"
\cite{guiselin} in which
chains are extremely polydisperse but, nonetheless, strongly extended
from the surface: the density of grafted
material decays with a power law of height, $\rhog(z) \propto z^{-2/3}$,
and the spacing between chains at height $z$ varies as
$\xi\sim z^{1/2}$.
We have tested these predictions by Monte Carlo simulation using
the bond-fluctuation model (BFM) \cite{bfm}.
Good agreement is obtained in $d=3$, which remains the upper
critical dimension for any $\nu$
(as may be confirmed by a simple reworking of Krug's argument \cite{krug}).

\begin{figure}
\centerline{\epsfig{file=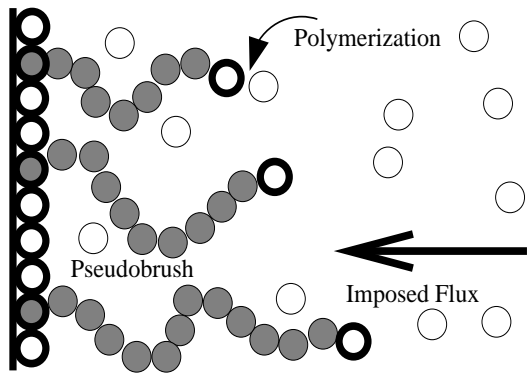,width=70mm,height=50mm,angle=0}}
\caption[]{Sketch of diffusive growth of a polymer layer by {\em in situ} 
polymerization.}
\label{figsketch}
\end{figure}

The continuum field description of the growth can be formulated quite
simply.
We define three coarse-grained densities (each expressed
as a local volume fraction): (i)
that of the aggregated monomers $\rhog(z,t)$; (ii) that of
active end monomers $\rhoe(z,t)$; and (iii) that of free
monomers $u(z,t)$.
These are coupled by three differential equations,
as follows. Firstly, we suppose
the polymerization is a local second-order process:
\begin{equation}
\partial_{t} \rhog = k \, \uf \, \rhoe
\label{eqsecondorder}
\end{equation}
where $k$ is a reaction constant.
Secondly, conservation of the particles implies:
\begin{equation}
\partial_{t} \rhog = - \partial_{t} \uf + D \Delta \uf = D \Delta \uf
\label{eqconservation}
\end{equation}
where $D$ is the diffusivity of free monomers.
The second equality embodies the ``adiabatic'' approximation,
valid in the quasi-static limit of infinitesimal flux \cite{wittensander}.
Thirdly we have a ``constitutive relation" between
local monomer density $\rhog$ and blob size, or equivalently
between $\rhog$ and the
end density, $\rhoe$:
\begin{equation}
\rhoe = - l \, \partial_{z} \rhog^{1/\as}
\label{eqphirho}
\end{equation}
Here $l$ is of order of the monomer size, $b$.
This relation arises because the number of chains ${\cal N}(z)$
per unit area of surface surviving to height $z$ is
(by simple geometry) of order $\xi(z)^{1-d}
\simeq \phi^{(1-d)/(1/\nu-d)}$, and the end density is
$-\partial {\cal N}/\partial z$.
Hence $\as$ obeys
$\as = (d-1/\nu)/(d-1)$.
Within this continuum approach, the exponent
$\as$ contains all required information about local chain structure.
Adopting Flory's estimate,
$\nu = 3/(2+d)$,  we have
$\as=2/3$ (which is, by accident, independent of $d$).

Next we consider the boundary conditions.
Within the continuum model,
as we show below, the density of adsorbing chain ends
is (formally) divergent near the wall, so the chance
of an incident particle actually reaching the wall is negligible.
We thus require that the density of the free monomers,
and its gradient, vanish there:
\begin{equation}
\uf(0,t)=\partial\uf(0,t)/\partial z=0.
\label{eqbyu1}
\end{equation}
Secondly, the boundary condition on $u$ at infinity is that appropriate to
a (small) constant flux $\ji$ of incoming monomers (each of volume $b^d$):
\begin{equation}
\lim_{z \rightarrow \infty} \partial \uf(z,t)/\partial z = b^{d} \ji/D
 \equiv \gi.
\label{eqbyu2}
\end{equation}
This ensures (in the adiabatic limit)
that the mass of the layer grows linearly in time.
For convenience, we may set $1/k = l =1$ by choosing
these as the units of time and length.
Two parameters then remain, the reduced flux $\gi$ and the
reduced diffusivity
$D/k l^2 = D$. The latter may itself
be set to unity by the rescalings
$\phi =
D^{\as} \tilde{\phi}, \rhoe = D \tilde{\rhoe}, u = D^{\as-1} \tilde{u}$
and $\gi = D^{\as-1}\tilde{\gi}$. In what follows we suppress the
tildes (i.e. we choose $D=1$) but
restore the scale factors later as required.

We now define a characteristic layer height by
the first moment \cite{moment} $\hc(t) =\int z\phi\,dz/\int \phi\, dz$,
and seek solutions of Eqs.(\ref{eqsecondorder}-\ref{eqbyu2})
in terms of the scaled variable
$x=z/\hc(t)$:
\begin{equation}
\rhog  =  z^{-\alpha} f(x)\;\;;\;\;\rhoe =  z^{-\beta} h(x) \;\;;\;\;
\uf =  z^{\gamma} g(x)
\label{eqansatz}
\end{equation}
Consistency of this scaling ansatz with the governing equations and
the boundary conditions fixes the exponents $\alpha,\beta,\gamma$ as
we now show.
Firstly, we eliminate $\partial_t\phi$ from
Eqs.~(\ref{eqsecondorder},~\ref{eqconservation})
to obtain $ \uf\rhoe = \partial_z^2 \uf $.
This forces the
end-density to decrease with an exponent $\beta =2$.
Secondly, the constant flux boundary conditions,
eqns.~(\ref{eqbyu1},\ref{eqbyu2}) impose
an exponent $\gamma=1$  (with $g(\infty) = \gi$ and $g(0)=0$).
Finally, we require on physical grounds that the function
$f(x)$ approaches a finite limit
at the origin, and that $f(\infty) = 0$.
The constitutive relation eq.~(\ref{eqphirho}), between monomer
and absorber density then reads
$
\as z^{-\beta}h(x) = z^{-(1+\alpha/\as)}f(x)^{1/\as}
\left({\alpha} - {x}(\ln f(x))' \right)
$
where the prime denotes a derivative.
Equating powers of $z$ we obtain
$ \beta = 1+\alpha/\as $;
within mean field theory
($\beta = 2$), this implies $\alpha = \as$
which fixes our final exponent.
It follows immediately that
$\xi \propto z^{1/(d-1)}$, and hence $\xi \propto \sqrt{z}$ in $3$
dimensions; this result is independent of $\as$.
In contrast, the total mass
$M = \int_0^\infty \phi(z)dz$ obeys
$M=\eta\hc^{1-\as}$ where $\eta(\as) = \int_0^\infty x^{-\as}f(x)dx$.
Under constant flux conditions,
$M=\gi t$, and one has accordingly $\hc(t)=(\gi t/\eta)^{1/(1-\as)}$.
This completes, at the scaling level, the generalization of the
mean field theory for needle
growth ($\as=1$) \cite{cates} to the case of arbitrary
$\as$. Note that, at this level, the only difference between the needle
model and the case of general $\as$ lies in the fact that, when a particle
is adsorbed, the adsorbing end moves forward by one monomer size for needle
growth but less than this for the polymer case. Since only the active ends
participate in screening, the scaling $\rho\sim z^{-2}$ remains
as for needles, and $\beta = 1+\alpha/\epsilon$ then follows from Eq.3.

To test these predictions, we have performed detailed
bond-fluctuation model simulations (for the method, see \cite{bfm}).
In this athermal model, ``local jumps" are made by choosing one monomer
at random (regardless of whether or not it is part of a chain)
and attempting to jump one lattice unit in a randomly chosen
basis direction.
The attempt is accepted if excluded volume and chain connectivity
requirements are satisfied.
If a free monomer happens to touch the wall at $z=a$
($a$ is the lattice spacing)
it adsorbs and becomes reactive.
A free monomer can polymerize at a reactive chain end
whenever the difference vector of the
two monomers is an allowed intrachain bond.
The simulation box is of size $L_{x}=L_{y}=100 a,L_{z}=500 a$
with periodic boundary conditions in $x-$ and $y$-directions and impenetrable
walls at $z=a$ and $z=L_{z}$.
The flux used was
$\ji = 1.3 \cdot 10^{-7}/\tau a^2$
where $\tau$ is the time for one Monte Carlo step per monomer.
No hydrodynamic interactions are taken into account.
In the adiabatic limit of interest, chains are fully relaxed
and the precise dynamics should be irrelevant.

\begin{figure}
\centerline{\epsfig{file=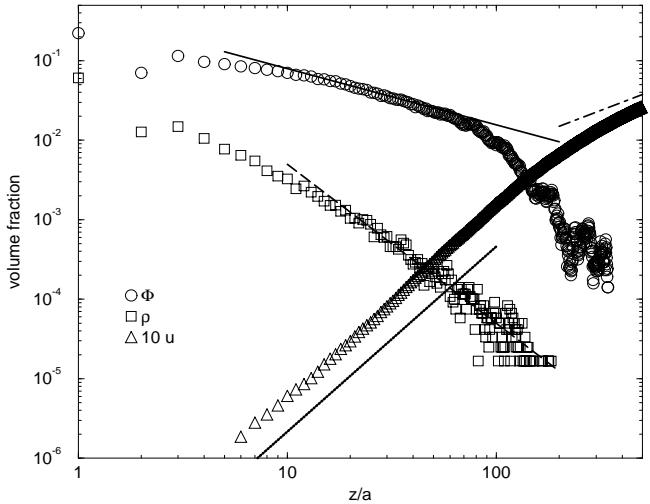,width=80mm,height=100mm,angle=-90}}
\caption[]{
The density profiles for the grafted monomers $\rhog$, the end monomers
$\rhoe$ and the free monomers $\uf$ versus distance $z$ from the wall.
An incident flux of $\ji = 1.3 \cdot 10^{-7}/\tau a^{2}$ was imposed over
a period of $4 \cdot 10^{6} \tau$.
We compare the densities with the power law predictions $\alpha = \as =2/3$
(solid line), $\beta=2$ (dashed line) and $\gamma=1$ (dashed-dotted line).
Near the wall the density of free monomers increases with distance
as $u(z) = z g(x) \sim z^{3-\as}$ (dotted line)
corroborating $\gamma=1$ and the asymptotic
power law solution of eq.(\ref{eqfxgx}) for small $x$ \cite{uslongpaper}.
The free monomer density (calculated from the incoming free monomer flux)
is very small and is vertically shifted.
}
\label{figphirhou}
\end{figure}

\begin{figure}
\centerline{\epsfig{file=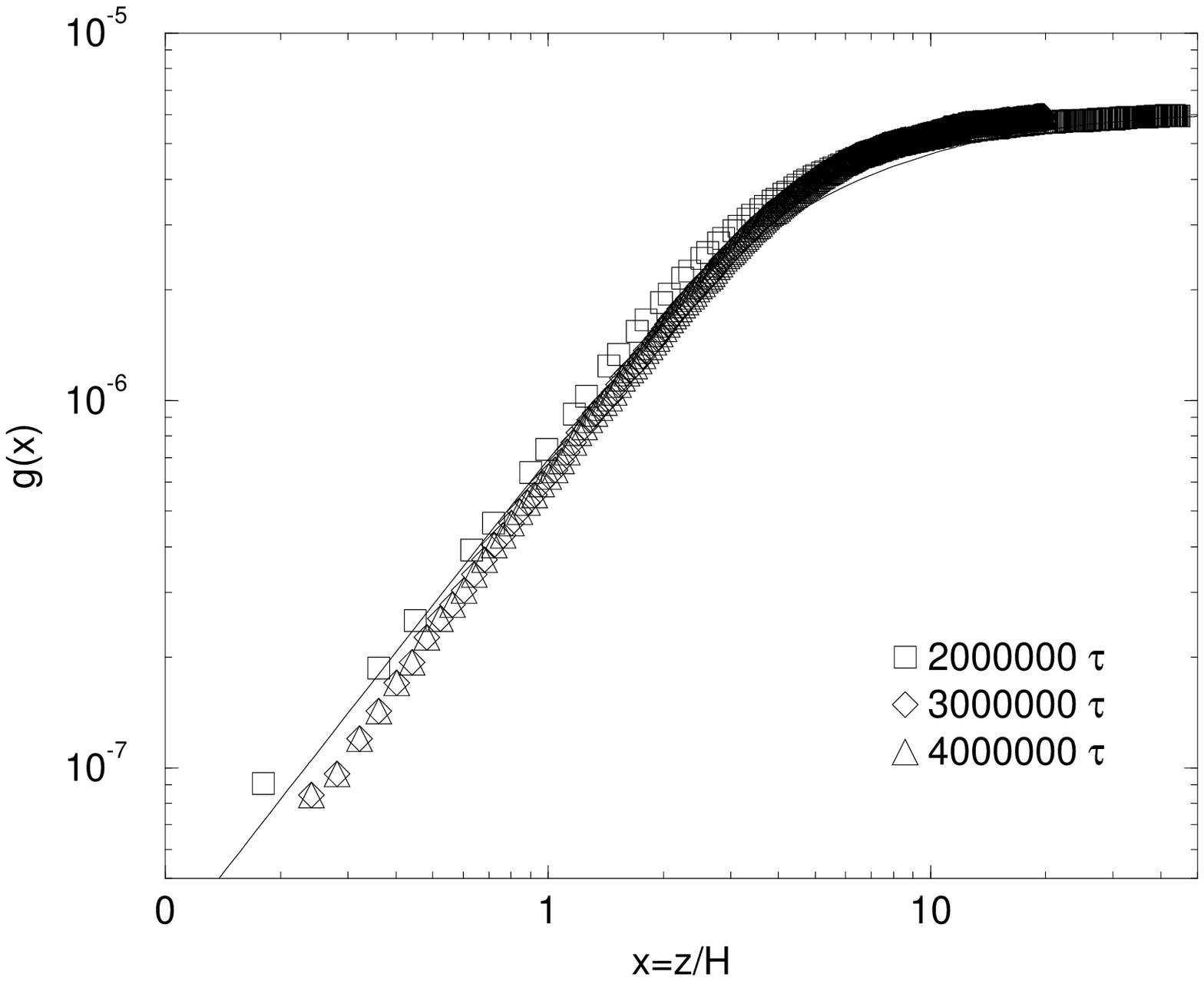,width=90mm,height=70mm,angle=0}
            \epsfig{file=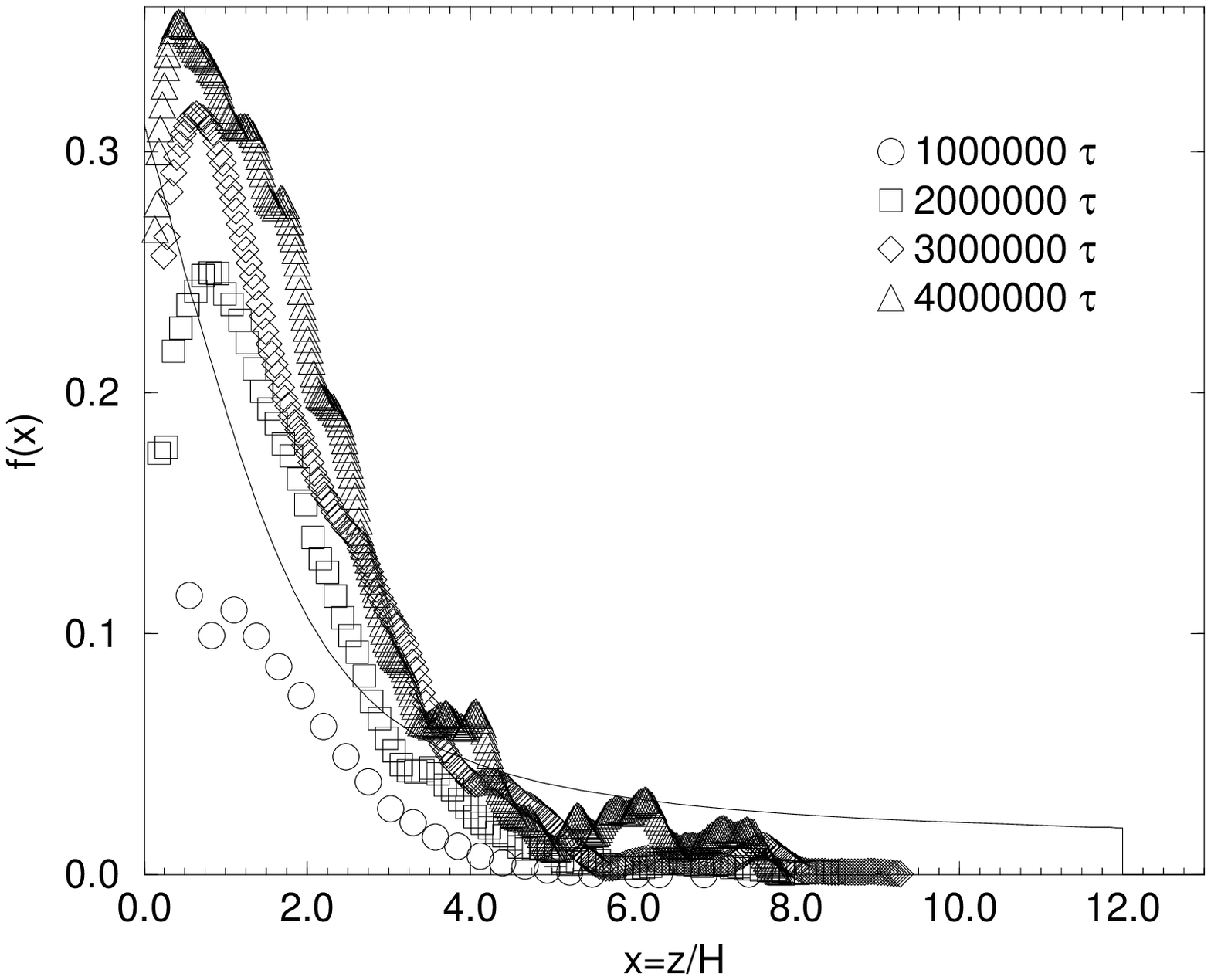,width=90mm,height=70mm,angle=0}}
\caption[]{
The scaling functions (a) $g(x)$ and (b) $f(x)$ for serveral times.
The solid line indicates the numerical solution of eq.~(\ref{eqfxgx}),
with a discontinuity at $x_{c}=12$, obtained by a Runge-Kutta method.
(a) The data collapse justifies the scaling
ansatz eq.~(\ref{eqansatz}) with an exponent $\gamma=1$.
(b) For $t = 10^{6} \tau$ the grafted chains are not stretched.
As the layer becomes more brush-like the curves merge.
}
\label{figgxfx}
\end{figure}

The density profiles for the grafted monomers $\rhog$, the end monomers
$\rhoe$ and the free monomers $\uf$ are shown in fig.~\ref{figphirhou}
and confirm broadly the scaling predictions.
In fig.~\ref{figgxfx} the scaling functions
$g(x)$ and $f(x)$ are shown.
The direct verification of the scaling prediction
for the monomer density is less successful than
for the free monomer density.
The accuracy of $f(x)$ is
roughly indicated by the size of the wiggles, which
apparently arise from statistical noise:
the accuracy decreases with both $x$ and $t$,
as the density becomes very low.
Also, the interactions of the first blobs with the wall
and fluctuations near the outer brush-edge,
omitted from our course-grained description,
imply two additional length scales
(the sizes of the first and the last blob of the layer) which,
judging from fig.~\ref{figgxfx} (b),
are non-negligible even at our longest times.

We now return to
the scaling functions $f(x)$, $g(x)$ and $h(x)$.
Defining a dimensionless parameter
$ \mu = {\gi\over(1-\as)\eta}$
the mean-field equations
~(\ref{eqsecondorder}-\ref{eqbyu2})
can be
reduced to the form:
\begin{eqnarray}
h(x)\, g(x) = - \mu x^{2-\as}f' &  = &
                 g(x) f(x)^{1/\as}
\left(1 - \frac{x f'(x)}{\as f(x)} \right) \nonumber \\
&  = & \, x \left( 2 g'(x) + x g''(x) \right)
\label{eqfxgx}
\end{eqnarray}
These equations (combined with the boundary conditions)
determine in principle the scaling functions
$f,g,h$; they
represent a powerful extension, to arbitrary $\as <1$, of
the existing results for needle growth ($\as =1$).
For the latter case, an exact solution of the
equations has in fact been
obtained \cite{cates}.
For $\as <1$, however,  we have found numerically (and
shown analytically \cite{uslongpaper}) that no acceptable
solutions satisfy the boundary conditions on $f$.
This suggests that either the scaling ansatz is wrong
(which seems unlikely, in view of the simulation data shown above),
or that new physics intervenes.
The latter may certainly be true within the outermost blob of
the polymer layer, in which both the mean-field averaging, and
the constitutive relation (3), must break down. Therefore we
postulate the existence of a step discontinuity in $f(x)$
at some value $x_c$, with $f=0$ for $x>x_c$.
(This sharp step, and corresponding features in $h$ and $g$, would in
reality be spread over the a scale
$\sim \hc(t)^{1/2}$, the size of an outermost blob.)
This ansatz is enough to
determine the scaling functions $f,g,h$ uniquely, including the value of
$x_c$ \cite{uslongpaper},
although in practice it is hard to obtain an accurate numerical
value; our best estimate is $x_c = 12 \pm 2$.
Numerical results from eq.~(\ref{eqfxgx}) for $f(x)$ and $g(x)$
are shown alongside the simulation data in figs.~\ref{figgxfx}.
Since $\hc(t)$ can be extracted directly from the data,
there is no ambiguity on the horizontal scale, but
the parameter values ($k,l$...) are not exactly known
for the simulation, so $g(x)$ is normalized vertically
by the plateau value at $x=\infty$.
On this basis, agreement is good for $g(x)$ and reasonable for $f(x)$.
The vertical scaling could be avoided by estimating
parameter values in the
simulation; agreement remains acceptable
\cite{uslongpaper}.

We now discuss the validity of the  adiabatic approximation
($\partial_{t} u \ll D \Delta u$)
used in eqn.~(\ref{eqconservation}) (see \cite{kassner}). By substituting the
scaling results derived above (restoring dimensional factors),
we find that this
assumption is selfconsistent for short times, but breaks
down when the layer reaches a crossover height
$h^{*} \approx (D/kl)^{\as/(1+\as)}\gi^{-(1+\as)}$.
This can be made arbitrarily large by
taking the limit of small incident flux and/or rapid diffusion.
For experimental purposes, it may anyway be more realistic to
impose, instead of a constant flux, an
asymptotic density of the free monomers in the bulk.
This would entail a second characteristic length (the depletion
length $\xi_D \sim\sqrt{Dt}$ of the incident diffusers).
So long as $h \ll \xi_D$, the
incident flux is $\ji \propto \ui \sqrt{D/t}$,
and the layer mass $M$ increases as
$\sqrt{t}$ rather than linearly. Expressed in terms of $M$,
however, the structural scaling exponents $\alpha$ and $\beta$
should be unaffected. Once again, the theory will break
down at long times.

We have shown how the case of diffusion-limited in-situ polymerization
of a polymer brush can be viewed as one of a large class of problems,
related to needle-growth models \cite{rossi,meakin,cates,krug},
each characterised by an exponent $\as$; $\as\simeq 2/3$
for polymerization in a good solvent whereas $\as=1$ recovers the needle
growth limit.
We found that the scaling
exponents for the density, end density and free
monomer density
obey $\alpha=\as,\beta=2,\gamma=1$.
These scaling predictions are
consistent with detailed Monte-Carlo
simulations based on the bond-fluctuation model. This is not obvious
because of the coarse-graining implicit the
continuum description -- which appears to be especially
delicate in the outermost blobs of the layer ({\em cf.} our discussion
of $f(x)$ near $x=x_c$).
An important extension of our work
is to the case of polymerization onto a spherical colloidal
particle \cite{vidal}.
We intend to discuss this in a future paper, along with results for
flat layers in $d=2$ \cite{uslongpaper}.

\vspace{0.3cm}

We are indebted to R.~Lipowsky and A.~Vidal for discussions.
This work was funded in part by EPSRC.


\end{document}